# CARVE: A glycan-specific framework for assessing protein–glycan binding-pose quality

Yifei Xiao[1], Huifeng Zhao[1], Silong Zhai[1], Zichang Jin[1], Tingjun Hou[1, *], Yu Kang[1, *]

[1]College of Pharmaceutical Sciences, Zhejiang University, Hangzhou 310058, Zhejiang, China

## Corresponding authors

**Yu Kang**

E-mail: yukang@zju.edu.cn

**Tingjun Hou**

E-mail: tingjunhou@zju.edu.cn

## Abstract

Protein–glycan interactions depend on the precise three-dimensional arrangement of carbohydrates at protein-binding interfaces, yet reliable assessment of predicted complex structures remains challenging. Existing evaluations commonly rely on fixed root-mean-square deviation (RMSD) thresholds, which do not fully account for the structural diversity and conformational flexibility of glycans. Here we developed CARVE, a glycan-specific framework for evaluating protein–glycan structure predictions. Through topology-aware glycan parsing, CARVE integrates orientation RMSD (oRMSD), interface-ring RMSD (irRMSD) and the fraction of native contacts (Fnat) to assess glycan scaffold geometry, interface-ring placement and native-contact recovery. These descriptors define four quality categories (CARVE-Level) with geometric thresholds that scale with the degree of polymerization (DP), and a continuous CARVE-Score, which reproduced those categories on 25,068 independently generated poses without recalibration (macro-F1 = 0.892). Pairing these measures with glycan-specific topological, chemical and geometric validity checks, we benchmarked nine prediction methods on CARVE-Bench, comprising 1,974 experimentally determined protein–glycan complexes. Among the top 25 predictions per target, HADDOCK3 achieved the highest Acceptable-level success rate (47.0%), whereas no method produced a high-quality, structurally valid pose for more than one in five. Collectively, CARVE provides a standardized basis for assessing protein–glycan structure prediction methods and guiding the development of more accurate docking and co-folding approaches.

## Introduction

Protein–glycan interactions mediate diverse biological processes, including cell recognition, immune regulation and host–pathogen interactions[1–4]. Understanding their molecular specificity requires structural information about how monosaccharide rings and their functional groups are presented at protein-binding interfaces[5–8]. Yet the diversity of glycans in stereochemistry, glycosidic connectivity and branching, together with their conformational flexibility, complicates experimental determination of protein–glycan complex structures[9–11]. Computational prediction of protein–glycan complex structures therefore offers an important complement to experimental approaches for investigating the molecular basis of carbohydrate recognition[10,12–15].

Computational approaches to protein–glycan complexes have developed from conventional molecular docking to carbohydrate-specific protocols and, more recently, deep learning methods[16–24]. Conventional docking methods such as AutoDock Vina (Vina) use conformational search and empirical scoring to position ligands in protein-binding sites[16]. Carbohydrate-specific approaches account for aspects of glycan flexibility: Vina-Carb incorporates Carbohydrate Intrinsic (CHI) energy corrections for glycosidic torsional preferences[19], whereas GlycanDock refines glycan poses through rigid-body and torsional sampling coupled to energy minimization[21]. HADDOCK3 uses prior information about the binding interface to guide docking[20]. More recently, deep learning methods have broadened the range of molecular complexes accessible to structure prediction. DiffDock[22] and CarsiDock were developed for protein–ligand docking, whereas AlphaFold 3[23] and Boltz-2 enable the joint prediction of biomolecular complex structures. Although not specifically developed for carbohydrates, their molecular representations offer opportunities for modelling glycan-containing complexes[26]. These approaches differ in their requirements for receptor structures, binding-site information and initial ligand conformations, as well as in their strategies for exploring and evaluating candidate structures.

Despite these advances, establishing a reliable picture of the capabilities and limitations of protein–glycan prediction methods remains challenging. Vina-Carb was compared with Vina on 101 protein–glycan complexes[19], GlycanDock evaluated local pose refinement on 109 complexes,[21] and HADDOCK3 assessed binding-site-informed docking on 89 complexes[20]. These studies provided valuable assessments of individual approaches, but their starting conditions and evaluation criteria differed, with measures ranging from pose and shape RMSD to ring-RMSD and interface-ligand RMSD. More recently, ProtGlycanDock compared ten methods using protein- and glycan-holdout test sets comprising 148 and 134 complexes, respectively[27]. Although this broadened method coverage, benchmark-specific holdout partitions do not necessarily ensure independence from the training data of pretrained models. Comparisons are further complicated by differences in receptor inputs, binding-site information and definitions of a successful prediction. More fundamentally, structural accuracy has frequently been assessed using fixed RMSD thresholds, sometimes alongside contact-based or composite measures[19,27,28], without a well-established basis for interpreting quality consistently across diverse glycan structures. Thus, reliable comparison requires not only representative test sets and standardized evaluation conditions, but also a defensible definition of what constitutes an accurate protein–glycan binding pose.

The difficulty of defining such standards is rooted in the molecular architecture of glycans and their recognition by proteins[29,30]. Glycosidic linkages connect stereochemically defined rings into flexible linear or branched structures, in which individual monosaccharide units may contribute differently to the binding interface. Recognition depends on the presentation of particular rings and functional groups and on the interactions they form with the receptor[6,31], while other portions of the glycan may remain more conformationally mobile. A geometric deviation may therefore reflect displacement of groups directly involved in recognition, rearrangement of peripheral regions, or both. An overall RMSD condenses these differences into a single value without identifying which aspects of the interface are retained[20]. Defining binding-pose quality consequently requires more than measuring structural deviations: it requires relating these measurements to the preservation of the native protein–glycan interface. Variation in glycan length and branching further complicates this interpretation, raising questions about whether common geometric tolerances can consistently represent comparable levels of structural quality. In protein–protein docking, frameworks such as CAPRI[32–34] and DockQ[35,36] have demonstrated how geometric accuracy and native-contact recovery can be integrated into interpretable quality categories and continuous scores. However, establishing analogous standards for protein–glycan complexes requires accounting for the distinct structural characteristics of carbohydrates. Furthermore, geometric agreement with an experimental reference does not necessarily ensure correct stereochemistry, anomeric configuration, glycosidic connectivity or ring geometry[37].

Therefore, a reliable assessment must link structural deviations to binding-interface fidelity while evaluating the validity of the glycan independently.

Here we introduce CARVE (Carbohydrate Assessment of Recognition, Validity and gEometry), a framework for defining protein–glycan binding-pose quality and systematically benchmarking structure prediction methods. CARVE establishes consistent structural comparisons through topology-aware glycan parsing and introduces three complementary descriptors: orientation RMSD (oRMSD), interface ring RMSD (irRMSD) and the fraction of native contacts (Fnat), which characterize complementary aspects of glycan positioning and native-contact recovery. Building on these descriptors, we establish degree of polymerization (DP)-dependent quality criteria (CARVE-Level) and develop CARVE-Score, a continuous quality measure calibrated against these criteria. We further construct CARVE-Bench, comprising 1,974 experimentally determined protein–glycan complexes, and evaluate nine prediction methods under consistent structural and glycan-specific validity criteria: the physics-based docking methods Glide SP[18], HADDOCK3, Vina-Carb, Vina and AutoDock-GPU[17]; the deep learning-based docking methods CarsiDock and DiffDock; and the co-folding models AlphaFold 3 and Boltz-2. Together, CARVE provides a structurally interpretable basis for defining protein–glycan prediction quality, comparing current prediction methods and assessing future advances.

## Glycan-specific descriptors distinguish scaffold and interface accuracy

Assessing the accuracy of protein–glycan binding poses requires consideration of both glycan conformation and the recovery of the native protein-binding interface, where the spatial arrangement of monosaccharide rings and hydroxyl groups governs molecular recognition[6,31,38]. A commonly used measure, ligand root-mean-square deviation (LRMSD)[20,36], quantifies the positional deviations of all glycan heavy atoms after alignment of the protein-binding interface. However, conformational flexibility around glycosidic linkages can produce substantial displacements in distal glycan regions without necessarily disrupting the binding interface. By assigning equal weight to all glycan heavy atoms, LRMSD conflates these distal deviations with errors in interface placement, without distinguishing where structural deviations occur or whether native protein–glycan contacts are preserved (Fig. 1a).

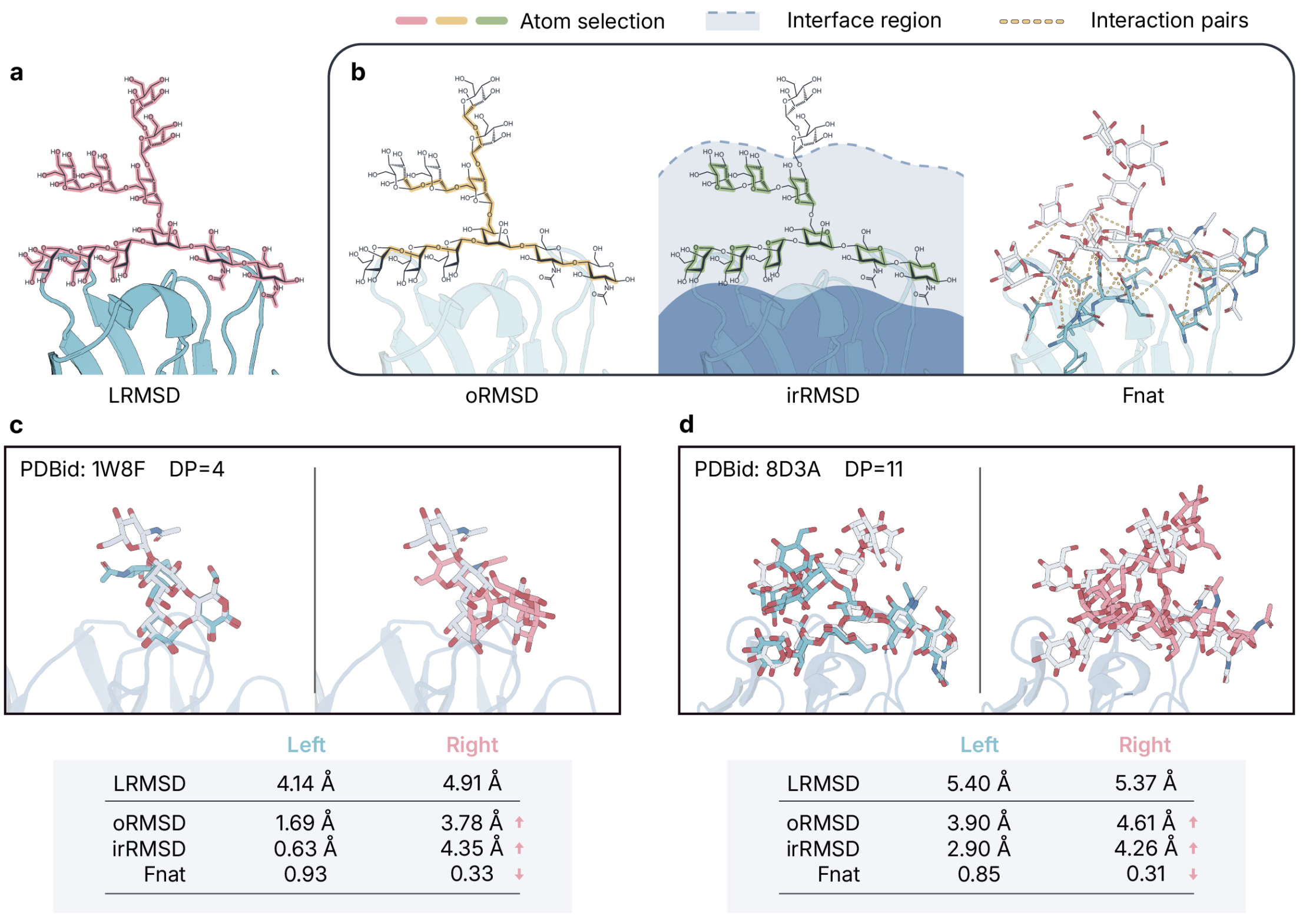

**Fig. 1 | Glycan-specific structural descriptors for evaluating protein–glycan binding-pose accuracy.**

**a,** Schematic illustration of conventional ligand root-mean-square deviation (LRMSD), which measures the overall displacement of glycan atoms (pink) relative to the reference structure after protein alignment. **b,** Schematic representations of the three complementary CARVE descriptors: oRMSD (left), irRMSD (middle) and Fnat (right). oRMSD evaluates the spatial arrangement of topology-defined glycosidic-scaffold atoms (yellow), whereas irRMSD focuses on ring atoms of reference-defined interface monosaccharide residues (green). Fnat quantifies the recovery of native protein–glycan contacts, illustrated by dashed lines. The interface region is highlighted in blue surface. **c,d,** Representative examples from PDB entries 1W8F (**c,** DP = 4) and 8D3A (**d,** DP = 11), illustrating differences in binding-pose accuracy that are not adequately reflected by LRMSD alone. For each complex, two predicted poses (cyan, left; pink, right) are superimposed on the experimental reference structure (grey). Despite comparable LRMSD values, the left poses exhibit lower oRMSD and irRMSD values and substantially higher Fnat, indicating more accurate glycan scaffold positioning, interface-ring placement and native-contact recovery.

To distinguish these structural contributions, we defined three glycan-specific descriptors based on monosaccharide identity and glycosidic connectivity (Fig. 1b). oRMSD quantifies positional deviations of topology-defined glycosidic-scaffold atoms across the glycan, providing a compact representation of its overall conformation and placement. irRMSD further focuses on ring atoms of reference-defined interface monosaccharide residues, identified using an 8.0 Å cutoff between the monosaccharide heavy-atom centroid and protein Cβ atoms (Cα for glycine), emphasizing their positional accuracy at the binding interface. To complement these geometric measures, Fnat quantifies the proportion of native protein-residue–monosaccharide contacts recovered in the predicted structure. Together, these descriptors provide complementary measures of glycan scaffold geometry, interface-ring placement and native-contact recovery.

The complementary information captured by these descriptors was evident in predicted poses with comparable LRMSDs but distinct structural deviations (Fig. 1c,d). In the tetrasaccharide complex 1W8F (degree of polymerization, DP = 4), two poses exhibited LRMSDs of 4.14 and 4.91 Å, yet the former closely reproduced all three interface rings (irRMSD = 0.63 Å; Fnat = 0.93), with deviations largely confined to a distal monosaccharide, whereas the latter showed substantial interface displacement (irRMSD = 4.35 Å; Fnat = 0.33). Their oRMSDs (1.69 versus 3.78 Å) further distinguished the extent of scaffold deviation. An even clearer discrepancy emerged in the branched glycan complex 8D3A (DP = 11), where nearly identical LRMSDs (5.40 and 5.37 Å) corresponded to markedly different interface-ring placement (irRMSD = 2.90 versus 4.26 Å) and native-contact recovery (Fnat = 0.85 versus 0.31). These observations illustrate that the magnitude of glycan displacement alone does not capture the distribution of structural errors or their consequences for native-contact preservation, highlighting the importance of evaluating these features separately.

## CARVE-Level establishes DP-dependent quality standards

Beyond distinguishing where structural deviations occur, assessing pose quality also requires considering how their magnitude relates to structural accuracy across glycan sizes (Fig. 2). In representative monosaccharide (DP = 1), disaccharide (DP = 2) and oligosaccharide (DP = 8) poses, comparable geometric deviations of approximately 1 Å corresponded to native-contact recovery (Fnat) values of 0.67, 0.93 and 0.82, respectively (Fig. 2d-f). This disparity became more pronounced at larger deviations (irRMSD = 1.7–1.9 Å; oRMSD = 1.5–1.7 Å), where Fnat decreased to 0.22 and 0.42 for the monosaccharide and disaccharide poses, respectively, but remained at 0.82 for the oligosaccharide (Fig. 2g-i). These differences indicate that a common geometric tolerance does not necessarily correspond to equivalent structural quality across glycan sizes, motivating DP-dependent quality criteria.

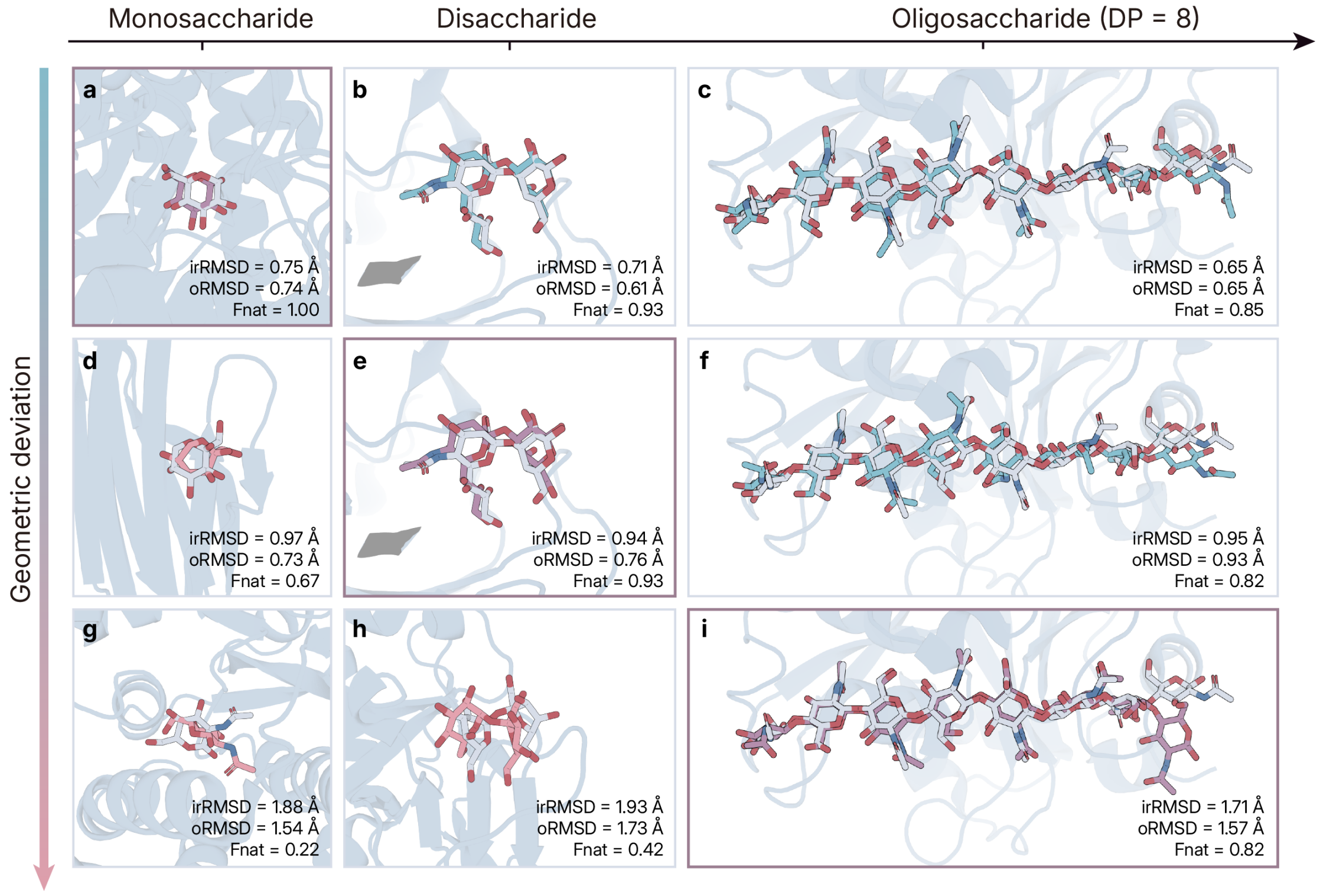

**Fig. 2 | Glycan length affects the structural significance of geometric deviation in protein–glycan binding poses.**

**a–i,** Representative protein–glycan binding poses illustrating the effects of glycan length on structural accuracy assessment. Examples are grouped into monosaccharides (**a,d,g**; DP = 1), disaccharides (**b,e,h**; DP = 2) and oligosaccharides (**c,f,i**; DP = 8), and arranged from top to bottom in order of increasing geometric deviation. Predicted glycan poses are superimposed on their experimental reference structures (grey). oRMSD, irRMSD and Fnat are indicated for each example.

These observations motivated the establishment of CARVE-Level, which classifies protein–glycan poses into four quality categories: High, Medium, Acceptable and Incorrect (Table 1). Building on the Critical Assessment of PRedicted Interactions (CAPRI) framework of combining geometric accuracy with native-contact recovery[32,39], we adopted uniform Fnat thresholds of 0.8, 0.5 and 0.2 for High, Medium and Acceptable quality, respectively, across all DP groups. In contrast, geometric thresholds for oRMSD and irRMSD were defined separately for DP = 1, DP = 2 and DP ≥ 3, with progressively stricter limits for shorter glycans, informed by ligand-size-dependent structural considerations and expert assessment of protein–glycan poses[20,40]. Each pose was assigned the highest quality level for which both geometric descriptors satisfied their DP-dependent thresholds and Fnat met the corresponding contact-recovery threshold.

**Table 1.** DP-dependent thresholds of CARVE-Level (H, High; M, Medium; A, Acceptable).

| DP group | Fnat $\geqslant$ (H / M / A) | irRMSD and oRMSD $\leqslant$ (H / M / A), Å |
|---|---|---|
| $DP = 1$ | 0.8 / 0.5 / 0.2 | 0.75 / 1.25 / 2.0 |
| $DP = 2$ | 0.8 / 0.5 / 0.2 | 1.0 / 1.5 / 3.0 |
| $DP \geq 3$ | 0.8 / 0.5 / 0.2 | 2.0 / 3.0 / 4.0 |

To assess whether these quality levels reflect chemically meaningful differences in protein–glycan recognition, we examined the recovery of native hydrogen bonds across the four CARVE-Level categories (Fig. 3). Across all three DP groups, mean hydrogen-bond recovery increased consistently from Incorrect to High, ranging from 0.14–0.27 for Incorrect poses to 0.75–0.89 for High poses, with intermediate recovery in the Acceptable and Medium categories. As hydrogen bonding depends on the spatial arrangement of interacting functional groups, this correspondence links CARVE-Level to chemically specific features of protein–glycan recognition beyond geometric deviation and general contact recovery.

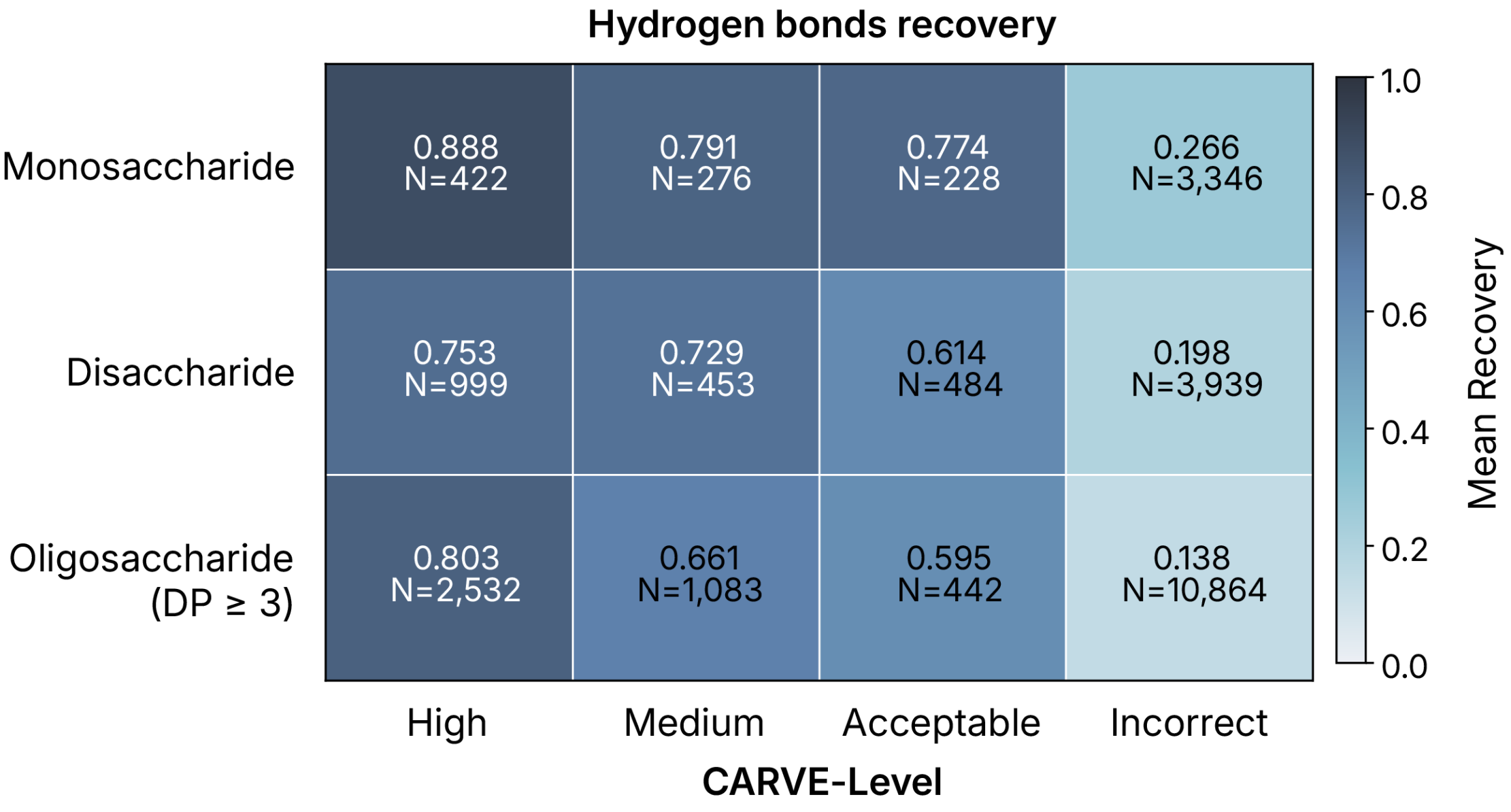

**Fig. 3 | Native hydrogen-bond recovery across CARVE-Level quality categories.**

Heatmap showing the mean fraction of native protein–glycan hydrogen bonds recovered by predicted binding poses across four CARVE-Level categories (High, Medium, Acceptable and Incorrect), stratified by glycan length: monosaccharides (DP = 1), disaccharides (DP = 2) and oligosaccharides (DP ≥ 3). Hydrogen-bond recovery was calculated as the fraction of hydrogen bonds present in the experimental reference structure that were preserved in the predicted pose. Values indicate the mean recovery within each group, with N denoting the number of evaluated poses.

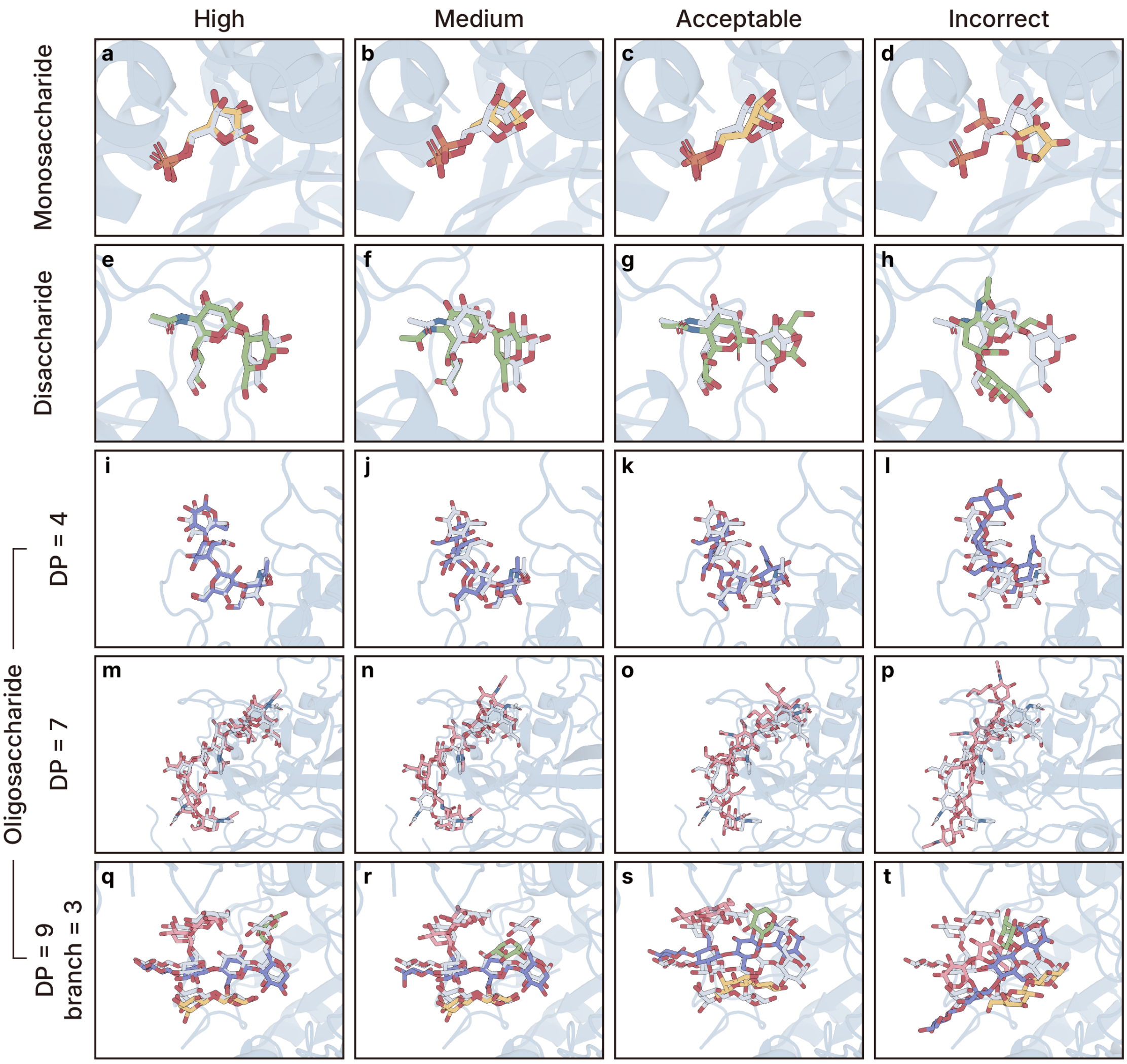

**Fig. 4 | Representative protein–glycan binding poses across CARVE-Level quality categories.** **a–t,** Representative protein–glycan binding poses classified into four CARVE-Level quality categories (High, Medium, Acceptable and Incorrect), illustrating structural differences across glycan sizes and branching architectures. Examples include a monosaccharide (**a–d**; DP = 1), a disaccharide (**e–h**; DP = 2) and oligosaccharides with DP = 4 (**i–l**), DP = 7 (**m–p**) and DP = 9 with three branches (**q–t**). Within each row, predicted glycan poses are superimposed on the corresponding experimental reference structure (grey). Columns represent progressively lower CARVE-Level quality categories from left to right, illustrating increasing deviations in glycan positioning and conformation relative to the experimental structures.

Representative structural overlays further illustrated the differences between quality levels across monosaccharides, disaccharides and oligosaccharides including branched glycans (Fig. 4). High-level poses closely reproduced the reference interface and glycan scaffold (Fig. 4a,e,i,m,q), whereas Medium (Fig. 4b,f,j,n,r) and Acceptable (Fig. 4c,g,k,o,s) poses retained aspects of the native binding mode despite progressively greater deviations in interface-ring placement and scaffold geometry. Incorrect poses (Fig. 4d,h,l,p,t) showed substantial departures from the reference binding geometry. Together, these observations support the structural and chemical interpretability of CARVE-Level across glycan sizes and architectures.

## CARVE-Score provides a transferable continuous quality scale

Building on the structurally interpretable quality standards established by CARVE-Level, we sought to develop a continuous measure that enables finer-grained assessment of predicted poses and quantitative comparisons across prediction methods. To this end, we developed CARVE-Score by normalizing oRMSD, irRMSD and Fnat against their corresponding CARVE-Level thresholds, transforming them into nonlinear similarity measures and combining the resulting values through harmonic aggregation to maintain sensitivity to errors in each structural component. To establish representative and non-redundant calibration datasets, we clustered glycan-binding sites by structural similarity and selected 750 diverse complexes for training and 75 complexes with low protein-structure similarity to the training set for independent validation. Training poses were generated from experimental structures by randomly perturbing dihedral angles around all rotatable glycan bonds, together with rigid-body rotations and translations, capturing both internal conformational and binding-placement deviations without introducing prediction method-specific biases. Balanced sampling across the four predefined quality categories yielded 900,000 poses for calibrating the scoring function. The resulting CARVE-Score preserved the established quality levels on a continuous scale, with score boundaries of 0.301, 0.532 and 0.817 delineating Incorrect, Acceptable, Medium and High quality (Fig. 5a).

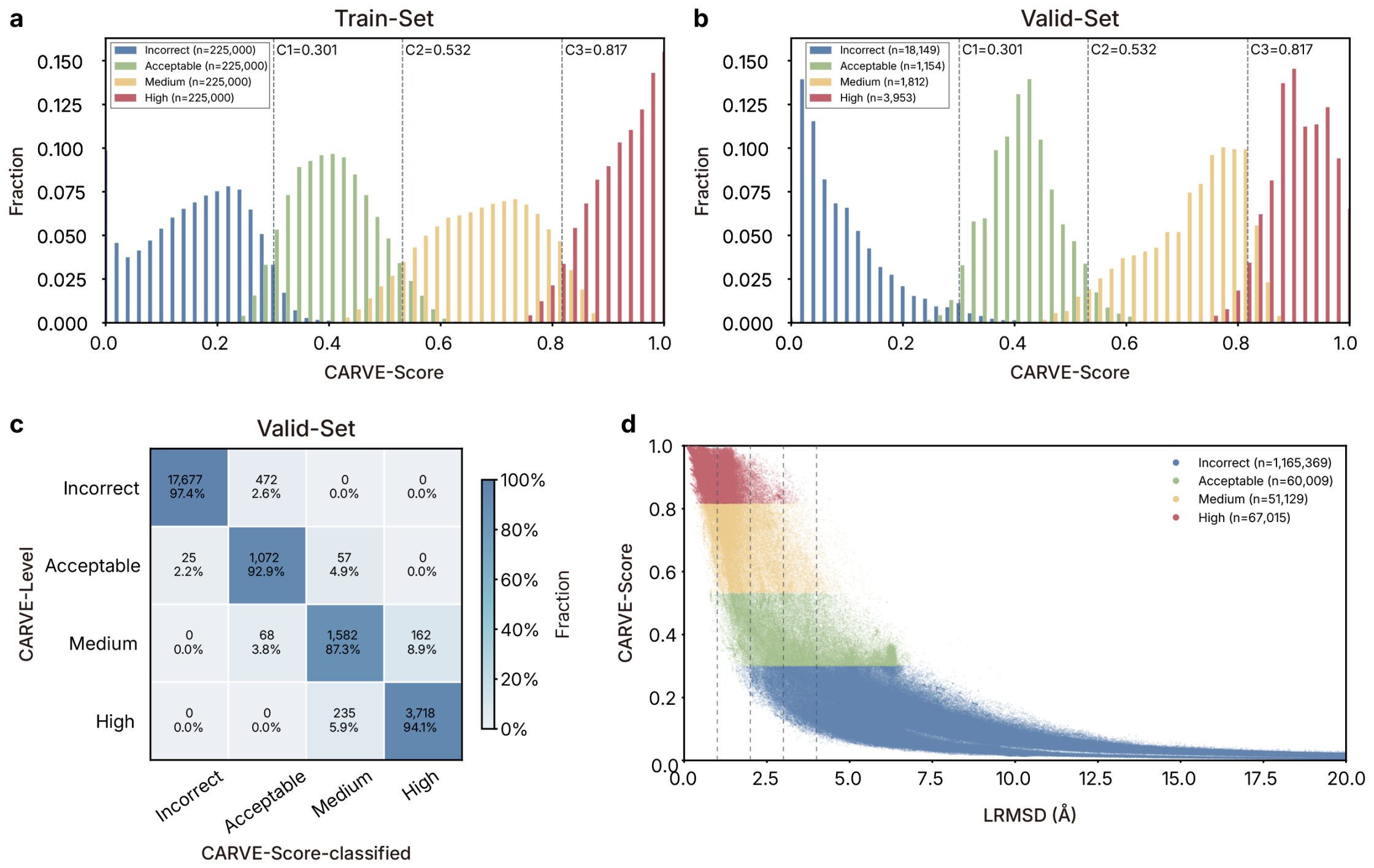

**Fig. 5 | Development and validation of CARVE-Score for continuous assessment of protein–glycan binding-pose quality.**

**a,b,** Distributions of CARVE-Score across four CARVE-Level quality categories (High, Medium, Acceptable and Incorrect) in the training (**a**) and validation (**b**) sets. Histograms show the fraction of poses within each CARVE-Level category across CARVE-Score intervals. Dashed vertical lines indicate the score thresholds (C1 = 0.301, C2 = 0.532 and C3 = 0.817) used to classify continuous scores into the four quality categories. **c,** Confusion matrix comparing CARVE-Level classifications with those assigned using CARVE-Score thresholds in the validation set. Each cell indicates the number of poses and the corresponding percentage within its CARVE-Level category, with color intensity representing the percentage. **d,** Relationship between CARVE-Score and conventional LRMSD across 1,343,522 predicted protein–glycan binding poses. Each point represents an individual pose, colored according to its CARVE-Level classification.

To assess whether this correspondence extends from controlled perturbations to actual model predictions, we applied the fixed scoring function and quality boundaries to 25,068 poses generated by four prediction methods for the 75 held-out complexes, without refitting. CARVE-Score maintained close agreement with the predefined CARVE-Level categories (Fig. 5b,c), achieving a macro-averaged F1 score (macro-F1) of 0.892, comparable to the training performance of 0.901. This agreement was consistent across DP groups and prediction methods, supporting the generalizability of CARVE-Score to protein–glycan complexes with low binding-site structural similarity to the training set and diverse prediction approaches without recalibration.

We next examined CARVE-Score across 1,343,522 poses from 1,974 protein–glycan complexes in the larger benchmark (Fig. 5d). Across this broader prediction set, CARVE-Score retained its correspondence with CARVE-Level, distinguishing differences in structural quality that were not resolved by LRMSD alone. These results establish CARVE-Score as a continuous, structurally interpretable measure for systematic evaluation of protein–glycan structure predictions.

## Systematic benchmarking of protein–glycan structure prediction

To systematically assess the current capabilities of protein–glycan structure prediction methods, we established CARVE-Bench comprising 1,974 experimentally determined protein–glycan complexes. The benchmark was temporally separated from the data used to develop and validate CARVE-Score, with a cutoff selected to minimize overlap with the structural training data of established prediction models, providing a more comparable basis for cross-method evaluation. We evaluated five physics-based docking methods, two deep learning-based docking methods and two co-folding approaches, with additional global docking configurations for Vina and Vina-Carb. Given the importance of preserving glycan chemical identity, stereochemistry, connectivity and molecular geometry, prediction success required both the corresponding CARVE-Score quality threshold and Pose-valid, combining topology-aware verification with PoseBusters-based geometry and placement assessments. The benchmark revealed a marked gap between the recovery of acceptable binding poses and the prediction of high-accuracy structures (Fig. 6a-c). Although HADDOCK3 achieved an Acceptable-level Top-25 success rate of 47.0%, its High-level success rate was only 11.2%. The relative ranking of methods also shifted with the required quality, with Vina-Carb leading at the Medium level (28.4%) and Boltz-2 and AlphaFold 3 achieving the highest High-level success rates (18.5% and 18.2%, respectively). Nevertheless, no method recovered a high-quality, structurally valid pose for more than one-fifth of the targets, even when considering 25 candidate predictions per complex.

We next examined how prediction performance varied across glycans of increasing DP (Fig. 6d). To ensure comparability, the analysis was restricted to the largest common set of protein–glycan complexes for which all evaluated methods produced analyzable poses. The mean CARVE-Score of structurally valid Top-25 poses declined markedly with increasing DP for several docking methods, particularly Glide SP and CarsiDock (Fig.

6d). In contrast, HADDOCK3 maintained comparatively high scores across a broad DP range, whereas co-folding approaches, such as AlphaFold 3, showed more stable performances, with higher scores for longer glycans. Vina-Carb and Vina performed similarly for shorter glycans, but Vina-Carb retained higher scores at DP ≥ 9.

Prior knowledge of the binding site also substantially influenced docking performance. For both Vina and Vina-Carb, restricting the search to the reference-defined binding region yielded higher success rates and a higher mean CARVE-Score than global docking without binding-site information, highlighting the additional challenge of identifying the correct glycan-binding region.

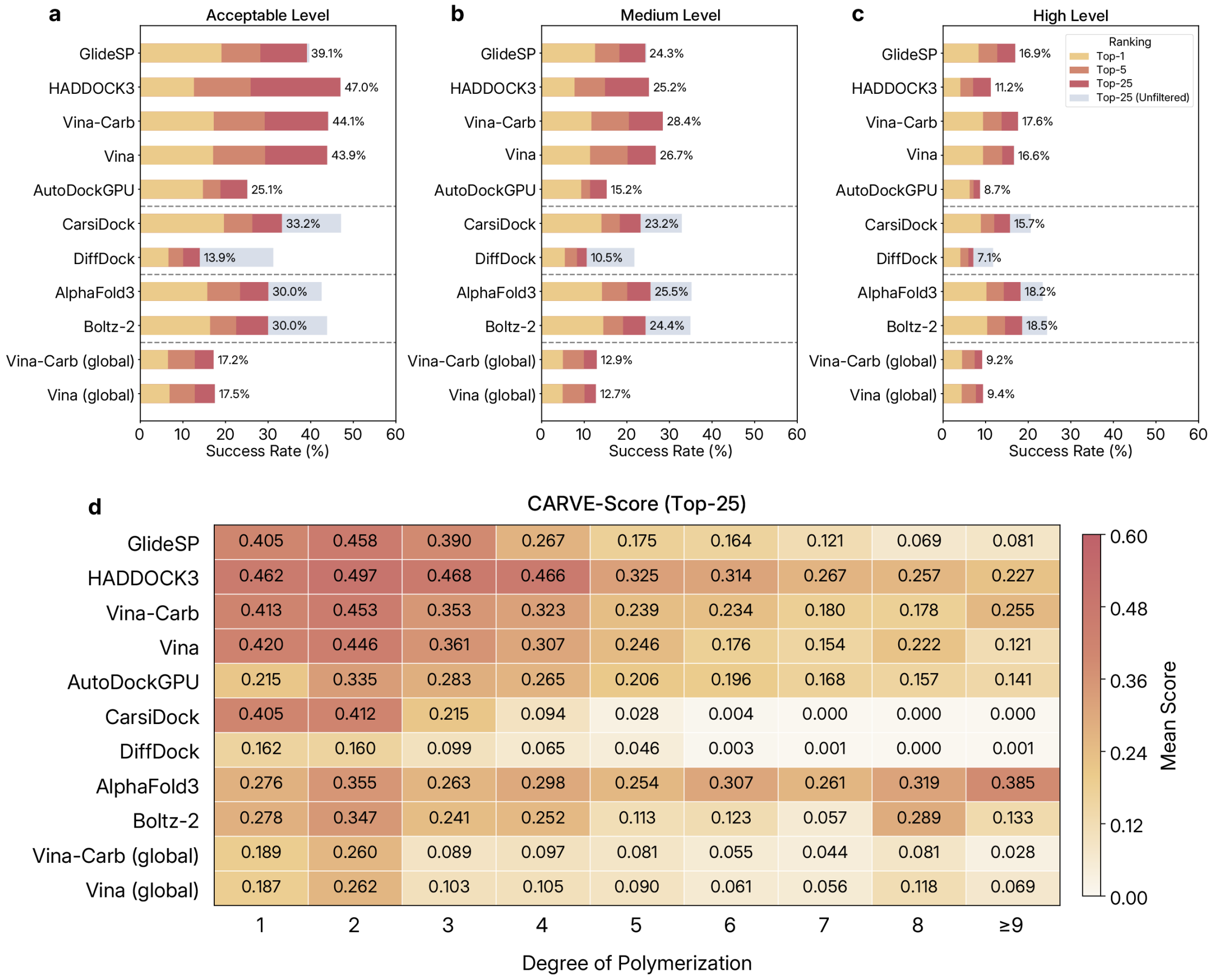

**Fig. 6 | Benchmarking protein–glycan binding-pose prediction methods across quality levels and glycan degree of polymerization.**

**a–c,** Prediction success rates of nine methods under eleven evaluation settings at the Acceptable (**a**), Medium (**b**) and High (**c**) CARVE-Level quality thresholds. Success rates are reported for the top-ranked (Top-1), top-five (Top-5) and top-25 (Top-25) predictions, with success defined as recovering at least one pose meeting or exceeding the corresponding quality threshold. The additional Top-25 (Unfiltered) bars indicate success rates before Pose-valid filtering. Global docking settings for Vina-Carb and Vina are shown separately. **d,** Heatmap showing the mean

CARVE-Score for the Top-25 predictions across glycan sizes, stratified by DP. Rows represent prediction methods and settings, and columns represent DP groups. Values indicate the mean CARVE-Score within each group, with color intensity reflecting score magnitude.

## Conclusion

Reliable assessment of protein–glycan binding poses requires more than measuring geometric deviations from experimental structures. The structural diversity and conformational flexibility of glycans demand evaluation criteria that capture distinct aspects of binding-pose accuracy. Here, we developed CARVE to provide a structurally interpretable framework for assessing protein–glycan binding poses. By integrating complementary descriptors of glycan orientation, interface-ring placement and native-contact recovery, CARVE enables a more comprehensive characterization of structural accuracy than conventional ligand RMSD alone. The DP-dependent CARVE-Level criteria translate these measurements into interpretable quality categories, whose structural relevance is supported by native hydrogen-bond recovery and representative structural analyses. CARVE-Score further provides a continuous measure of prediction quality while maintaining consistency with the established quality standards.

Applying this framework to 1,974 experimentally determined protein–glycan complexes from nine methods revealed substantial limitations in current prediction capabilities. Although the evaluated methods differed in their performance across quality levels, none achieved a Top-25 success rate exceeding 20% when both high-quality binding-pose accuracy and glycan-specific validity were required. These findings highlight the importance of considering structural quality and molecular validity together when interpreting prediction performance, rather than relying solely on conventional geometric success criteria.

CARVE integrates interpretable quality criteria, a continuous scoring function and systematic benchmarking to provide a unified foundation for evaluating diverse prediction approaches. Beyond comparing existing methods, this framework enables more consistent identification of structural inaccuracies and provides quantitative criteria for assessing methodological improvements. By establishing a clearer definition of prediction quality, CARVE lays the groundwork for more reliable evaluation and continued development of protein–glycan structure prediction methods.

## Methods

### Dataset collection and partitioning

#### *Collection and preparation of protein–glycan complexes*

Experimentally determined protein–glycan complexes were collected from the RCSB Protein Data Bank (PDB)[41,42], considering structures released before 25 October 2025. Complexes were retained if they contained a protein chain of at least 50 amino acid residues and a non-covalently bound glycan. Structures with missing protein atoms within 5.0 Å of the glycan were excluded to ensure structural completeness at the binding interface. The retained structures were processed using the Protein Preparation Wizard module in the Schrödinger suite[43], including removal of water molecules, assignment of bond orders, addition of hydrogen atoms, and reconstruction of missing protein side chains and loops using Prime.

#### *Temporal partitioning and CARVE-Bench construction*

To separate CARVE-Score development from subsequent model benchmarking and minimize potential overlap with the training data of established prediction methods, the collected complexes were partitioned by their PDB release dates using 30 September 2021 as the cutoff. The 17,006 complexes released on or before this date formed the candidate pool for CARVE-Score training and validation, whereas 5,896 post-cutoff complexes were reserved for construction of CARVE-Bench. From the post-cutoff set, one representative glycan-binding site was retained per PDB entry, yielding 1,974 protein–glycan complexes that constitute CARVE-Bench. This temporal separation ensures that the benchmark structures were excluded from the development and validation of CARVE-Score while reducing potential training-data overlap with established prediction methods. Because training and model-selection dates vary among prediction methods, the cutoff was not taken to establish that every benchmark structure was unseen by every method.

#### *Training and validation datasets for CARVE-Score*

To construct structurally diverse datasets while minimizing redundancy between training and validation, glycan-binding sites were extracted from the pre-cutoff complexes by selecting protein residues within 10 Å of any glycan heavy atom. Binding-site structural similarity was assessed using Foldseek[44,45], with a template modelling score (TM-score) threshold of 0.65 for clustering. Training and validation complexes were selected from non-overlapping structural clusters to minimize similarity between the two datasets.

To ensure representation across different glycan sizes, complexes were stratified by DP into three groups: DP = 1, DP = 2 and DP $\geqslant$ 3. The training set comprised 750 complexes, including 150 monosaccharide, 200 disaccharide and 400 oligosaccharide complexes. An additional 75 complexes, comprising 15, 20 and 40 complexes from the respective DP groups, were reserved for independent validation. Each validation complex belonged to a distinct structural cluster, with no clusters shared between the training and validation sets.

The training complexes served as experimental reference templates for generating controlled conformational perturbations used to calibrate CARVE-Score. In contrast, validation was performed using predictions generated by AlphaFold 3, Boltz-2, HADDOCK3 and Vina-Carb, with up to 100 poses generated per method and complex. The resulting validation dataset comprised 25,068 evaluable poses, providing an assessment of CARVE-Score on independently generated predictions.

## Glycan parsing and structural alignment

### *Glycan parsing and atom mapping*

A topology-aware glycan parser was developed to identify and standardize glycan structures independently of the residue and atom naming conventions in the input glycan files. The parser resolves monosaccharide identities, ring forms, stereochemical configurations, glycosidic connectivity and branching topology, establishing standardized residue and atom identifiers for subsequent structural comparison. The standardization procedure also accommodates recoverable structural inconsistencies encountered in predictions from AlphaFold 3 and Boltz-2, enabling chemically consistent atom mapping between reference and predicted glycans. Predictions with severe molecular topology alterations that could not be resolved to supported monosaccharide structures in the parser's topology library were classified as unscorable and excluded from subsequent geometric evaluation.

### *Receptor alignment*

To evaluate glycan placement relative to the protein receptor, predicted complexes were aligned to their experimental reference structures using the protein-binding region. Protein residues within 10 Å of any glycan heavy atom in the reference complex were first identified, and a continuous segment was defined for each corresponding protein chain, extending from the first to the last selected residue in sequence order. The reference and predicted protein segments were then superposed using their corresponding heavy atoms. All glycan RMSDs, including LRMSD, oRMSD and irRMSD, were calculated in this receptor-aligned coordinate frame, without additional superposition of the glycan.

### Structural descriptors and quality criteria

Structural deviations between reference and predicted glycans were quantified using the root-mean-square deviation (RMSD) of corresponding atoms. For a predefined set $A$ of corresponding atoms, RMSD was calculated as:

$$\mathrm{RMSD}(A) = \sqrt{\frac{1}{N_A}\sum_{i=1}^{N_A}\left|\mathbf{x}_i^{\mathrm{pred}} - \mathbf{x}_i^{\mathrm{ref}}\right|^2}$$

where $N_A$ is the number of paired atoms. LRMSD was calculated using all matched glycan heavy atoms, whereas the glycan-specific geometric descriptors, oRMSD and irRMSD, used chemically defined subsets of these atoms.

### *Glycan orientation RMSD (oRMSD)*

oRMSD quantifies deviations in overall glycan scaffold geometry using atoms selected according to monosaccharide identity and glycosidic connectivity. For each glycosidic linkage, the selected atoms comprise the carbon path from the anomeric carbon (C1 or C2) to the linkage-bearing carbon (Cp), together with the bridging oxygen (Op) and the donor carbon (Cd), forming the linkage segment Cp–Op–Cd. All intervening carbon atoms along the numbered ring path are included. The topology-defined scaffold atom set was determined from the reference glycan and applied consistently to its corresponding predicted structures.

The resulting topology-defined atom set $\mathrm{A_o}$ is used to calculate:

$$\mathrm{oRMSD} = \mathrm{RMSD}(\mathrm{A_o})$$

### *Interface-ring RMSD (irRMSD)*

irRMSD measures the positional deviation of monosaccharide rings at the protein-binding interface. Interface monosaccharides were identified from the reference complex using the distance between each monosaccharide's heavy-atom centroid and the Cβ atoms of protein residues (Cα for glycine). A monosaccharide was classified as interfacial if its centroid was within 8.0 Å of at least one of these protein atoms.

For each interface monosaccharide, all endocyclic ring atoms were selected according to its chemical identity. The same reference-defined interface monosaccharides and corresponding ring atoms were used for all predicted poses, ensuring a consistent evaluation region independent of the predicted binding geometry.

The resulting interface-ring atom set $A_{ir}$ was used to calculate:

$$\mathrm{irRMSD} = \mathrm{RMSD}(A_{ir})$$

### *Fraction of native contacts (Fnat)*

The fraction of native contacts (Fnat) measures the recovery of protein–glycan contacts observed in the experimental reference structure. Contacts were defined at the protein-residue–monosaccharide level. A protein residue and a monosaccharide were considered to be in contact if any pair of their heavy atoms was separated by less than 4.0 Å. Fnat was calculated as:

$$F_{nat} = \frac{|\mathcal{C}_{ref} \cap \mathcal{C}_{pred}|}{|\mathcal{C}_{ref}|}$$

where $\mathcal{C}_{ref}$ and $\mathcal{C}_{pred}$ denote the contact sets of the reference and predicted complexes, respectively. Fnat therefore ranges from 0 to 1, with higher values indicating greater recovery of native protein–glycan contacts.

### *CARVE-Level quality criteria*

CARVE-Level combines oRMSD, irRMSD and Fnat to classify protein–glycan poses into four structural quality categories: High, Medium, Acceptable and Incorrect. Following the CAPRI framework of combining geometric accuracy with native-contact recovery, the Fnat thresholds were set to 0.8, 0.5 and 0.2 for High, Medium and Acceptable quality, respectively. These thresholds were applied uniformly across all DP groups.

In contrast, geometric thresholds for oRMSD and irRMSD were defined separately for DP = 1, DP = 2 and DP ≥ 3, reflecting the different structural implications of comparable positional deviations across glycan sizes. The resulting thresholds are summarized in Table 1.

For each quality category, a pose was required to satisfy the corresponding oRMSD and irRMSD thresholds simultaneously, together with the minimum Fnat requirement. Each pose was assigned the highest category for which all three criteria were satisfied. Poses failing to meet the Acceptable criteria were classified as Incorrect.

***Assessment of native hydrogen-bond recovery***

To examine the chemical relevance of CARVE-Level, native hydrogen-bond recovery was evaluated by comparing hydrogen bonds in the predicted protein–glycan complexes with those identified in their experimental reference structures. Hydrogen bonds were identified using Protein-Ligand Interaction Profiler (PLIP)[46], and the recovery fraction for each pose was calculated as the proportion of reference hydrogen bonds reproduced in the prediction.

## CARVE-Score construction and calibration

CARVE-Score integrates oRMSD, irRMSD and Fnat into a continuous measure of protein–glycan pose quality while retaining the DP-dependent quality criteria established by CARVE-Level. The scoring function comprises three stages: normalization to CARVE-Level boundaries, nonlinear similarity transformation and harmonic aggregation. Its parameters were calibrated using level-balanced conformational decoys generated independently of any specific structure prediction method.

***Normalization to CARVE-Level boundaries***

To place the three metrics on a common scale, their errors were defined as $e_F = 1 - F_{nat}$, $e_{ir} = irRMSD$ and $e_o = oRMSD$. For each descriptor, the High, Medium and Acceptable error boundaries were denoted by $(h, m, a)$.

Each error was mapped to a dimensionless boundary distance $D$ using a continuous piecewise-linear transformation:

$$D(e;h,m,a)=\begin{cases}\frac{e}{h}, & 0\le e\le h,\\ 1+\frac{e-h}{m-h}, & h<e\le m,\\ 2+\frac{e-m}{a-m}, & e>m.\end{cases}$$

Under this transformation, $D=1$, $D=2$ and $D=3$ correspond to the High, Medium and Acceptable boundaries, respectively, regardless of glycan DP. The CARVE-Level classification can consequently be expressed as a maximum-distance rule:

$$L=\begin{cases}3, & D_{max}\le 1,\\ 2, & 1<D_{max}\le 2,\\ 1, & 2<D_{max}\le 3,\\ 0, & D_{max}>3,\end{cases}\qquad D_{max}=max(D_F,D_{ir},D_o)$$

where $L=3,2,1,0$ denote High, Medium, Acceptable and Incorrect, respectively.

***Nonlinear similarity transformation***

The normalized boundary distances were converted into bounded similarity values using a Lorentzian function:

$$s_k=\frac{1}{1+(D_k/\sigma_k)^2},\quad k\in\{F,ir,o\}$$

where $\sigma_k$ denotes the scale of each metric.

***Harmonic aggregation and score calibration***

The three similarity values were combined using an equal-weight harmonic mean to obtain the final CARVE-Score:

$$S=\frac{3}{\frac{1}{s_F}+\frac{1}{s_{ir}}+\frac{1}{s_o}}=\frac{3}{3+\left(\frac{D_F}{\sigma_F}\right)^2+\left(\frac{D_{ir}}{\sigma_{ir}}\right)^2+\left(\frac{D_o}{\sigma_o}\right)^2}$$

The harmonic mean places greater emphasis on lower component similarities, limiting the extent to which accurate recovery of one structural feature can compensate for substantial errors in another. The resulting CARVE-Score $S$ lies in the interval (0,1], with higher values indicating greater agreement with the reference structure under the combined quality criteria.

Three score cutoffs, $C_1$, $C_2$, $C_3$, were introduced to map the continuous score onto the four predefined CARVE-Level categories.

### *Calibration on level-balanced decoys*

To calibrate CARVE-Score without introducing the characteristic error distributions of individual prediction methods, synthetic poses were generated by controlled perturbation of experimental glycan structures from the 750 training complexes. Protein receptors were retained in their reference coordinates, while glycan structures were perturbed through rigid-body translations and rotations, together with torsional changes around glycosidic linkages and other rotatable glycan bonds, including exocyclic substituents. These perturbations introduced both internal conformational deviations and changes in glycan placement relative to the receptor.

For each candidate, a target quality category was selected from the least-populated category, and a noise variable $u$ was sampled uniformly from a predefined level-specific interval. The perturbation scale for each class $k$ was determined by geometric interpolation:

$$\delta_k(u) = \delta_{k,min}\left(\frac{\delta_{k,max}}{\delta_{k,min}}\right)^u$$

Candidate poses were screened for intramolecular steric clashes and assigned a CARVE-Level using the original geometric and contact-recovery criteria. Poses were retained according to their realized quality categories until 300 poses had been collected for each of the four levels per training complex. This procedure yielded 900,000 calibration poses, with equal representation of all four quality categories within every complex.

### *Parameter optimization and independent validation*

The scaling parameter $\sigma_F$ was fixed at 1.6, whereas $\sigma_o$, $\sigma_{ir}$ and the three score boundaries were optimized to maximize agreement between CARVE-Score-derived categories and the predefined CARVE-Level assignments.

Optimization followed a nested procedure. For each candidate pair $(\sigma_{ir}, \sigma_o)$, the three score boundaries were independently optimized for the cumulative binary classification tasks $L \geq \ell$ by maximizing the Matthews correlation coefficient (MCC):

$$C_\ell = \underset{c\in\mathcal{T}}{arg\,max}\; MCC\left(\mathbb{1}[L_j \geq \ell],\ \mathbb{1}[S_j \geq c]\right), \qquad \ell = 1,2,3$$

Here, $j$ indexes the calibration poses and $\mathcal{T}$ comprises all distinct candidate score values.

For each candidate pair of scaling parameters, the optimized boundaries were used to assign continuous-score categories. The scaling parameters were then selected to maximize the macro-averaged F1 score over the four CARVE-Level categories:

$$(\sigma_{ir}^{\star}, \sigma_{o}^{\star}) = \underset{\sigma_{ir},\sigma_{o}}{arg\,max}\ \frac{1}{4}\sum_{q=0}^{3} F1_q\left(L, \hat{L}\left(S;\ \sigma_{ir}, \sigma_{o}, C^{\star}(\sigma_{ir}, \sigma_{o})\right)\right)$$

An exhaustive 14 × 14 grid search was performed over the ratios $\sigma_{ir}/\sigma_F$ and $\sigma_o/\sigma_F$, with non-uniformly spaced values from 0.1 to 12. This was followed by a finer search with a step size of 0.1 within ±0.5 of the coarse optimum.

The parameters were $\sigma_F = 1.60$, $\sigma_o = 1.60$ and $\sigma_{ir} = 1.84$, with score boundaries $C_1 = 0.3009$, $C_2 = 0.5316$ and $C_3 = 0.8174$.

## Glycan-specific structural validity assessment

To assess the validity of predicted protein–glycan complexes, we established a glycan-specific validation framework that combines topology-aware glycan identification with molecular geometry and protein–glycan placement checks adapted from PoseBusters[28]. Structural validity was evaluated in three categories: glycan identity and topology, molecular geometry, and protein–glycan placement.

### *Glycan identity and topology consistency*

Glycan identity and topology were assessed using the topology-aware parser described above. The parser identifies constituent monosaccharide residues and their stereochemical configurations, ring forms, glycosidic linkage positions and branching relationships directly from the predicted molecular structure. These assignments were compared with the corresponding reference glycan to assess consistency in chemical identity, stereochemistry and molecular connectivity.

### *Molecular geometry and protein–glycan placement*

Molecular geometry was evaluated using PoseBusters-based checks of bond lengths, bond angles, intramolecular steric clashes and applicable ring and double-bond geometry constraints. These assessments were used to identify deviations from chemically plausible glycan geometries.

Protein–glycan placement was evaluated separately to identify physically implausible intermolecular arrangements, including severe steric clashes and excessive molecular overlap between the glycan and protein receptor. These checks assess the physical compatibility of the predicted glycan with its protein environment independently of its agreement with the experimental binding pose.

***Structural validity classification***

A predicted pose was classified as Pose-valid only if it passed all required checks of glycan identity and topology, molecular geometry, and protein–glycan placement. Failure of any required check resulted in a non-valid classification.

Structural validity was evaluated separately from pose accuracy. In CARVE-Bench, a prediction was considered successful at a given quality level only when it satisfied both the corresponding CARVE-Score threshold and all Pose-valid requirements.